\magnification=\magstep1
\baselineskip=17pt
\hfuzz=6pt

$ $

\vskip 1in

\centerline{\bf Decoherent histories approach to the cosmological
measure problem}

\bigskip

\centerline{Seth Lloyd}

\centerline{Department of Mechanical Engineering, Department of Physics}

\centerline{Massachusetts Institute of Technology}

\vskip 1cm

\noindent{\it Abstract:}  The method of decoherent histories allows 
probabilities to be assigned to sequences of quantum events in systems,
such as the universe as a whole, where there is no external observer
to make measurements.  This paper applies the method of decoherent 
histories to address cosmological questions.   Using a series of simple
examples, beginning with the harmonic oscillator, 
we show that systems in a stationary state 
such as an energy eigenstate or thermal state
can exhibit decoherent histories with non-trivial
dynamics.  We then examine decoherent histories in a universe that
undergoes eternal inflation.  Decoherent histories that
assign probabilities to sequences of events in the vicinity of a 
timelike geodesic supply a natural cosmological measure.   Under
reasonable conditions, such sequences of events do not suffer
from the presence of unlikely statistical fluctuations that mimic
reality.

\vskip 1cm

The consistent or decoherent histories approach to quantum mechanics 
is a method for assigning probabilities to sequences of events 
for a quantum-mechanical system [1-35].  Because it does not rely
on the notion of measurement, the decoherent histories approach is useful
in theories such as quantum cosmology where probabilities have to be
assigned, but no external system is measuring or decohering the system --
in this case, the universe itself.   The cosmological measure problem
addresses the question of how to assign probabilities to events in various
cosmological scenarios, e.g., eternal inflation [36-54].  
Some apparently reasonable
cosmological measures have counter-intuitive consequences, e.g., the notion
that we are all just `Boltzmann brains' that arose from a statistical 
fluctuation.  This paper applies the method of decoherent histories to 
the cosmological measure problem.  First, the formalism of decoherent 
histories is reviewed.  Second, the method is applied to simple systems
to show that stationary states such as
energy eigenstates or thermal states
can still exhibit non-trivial decoherent histories,
in contradiction to the natural intuition that systems in stationary
states `do nothing.'   (This result answers
a question raised by Boddy {\it et al.} about whether systems in
stationary states can exhibit decoherent histories that
fluctuate over time [54].)  Decoherent histories are used to show
that systems described by a stationary state of quantum jump models
or environmentally induced decoherence models can still be thought 
of as exhibiting non-trivial temporal fluctuations.
Third, decoherent histories are applied to eternal inflation. 
Decoherent histories for the sequences of events that occur in
the vicinity of a timelike geodesic are shown to give rise to 
cosmological measures that differ from conventional volume-counting measures.
Under conventional assumptions about the potentials in the underlying
physics, such histories give rise to a picture of eternal inflation
in which a period of rapid inflation gives rise
to a Friedman-Robertson-Walker cosmology in de Sitter space with
small cosmological term.   After a period long compared with the
Hubble time of the de Sitter space, but short compared with the time
required to generate a Boltzmann brain, quantum/thermal fluctuations
give rise to another period of rapid inflation, and the cycle recommences.

The decoherent histories approach originated with Griffiths [1-3], who
called this approach consistent histories, and was independently
developed by Omn\`es [4-7] and by Gell-Mann and Hartle [8-15], who termed this
approach decoherent histories.  We will adopt the latter nomenclature.
Consider a sequence of projective measurements that could be made
on a quantum system
at times $t_1 < \ldots < t_n$. (The extension to generalized measurements
will be given below.)  The measurement at time $t_k$
has exhaustive and mutually exclusive outcomes $\alpha_k$.  This
measurement corresponds to a set of projection operators
$\{P^k_{\alpha_k} \}$ in the Heisenberg picture, where
$P^k_{\alpha_k} P^k_{{\alpha'}_k} = \delta_{\alpha_k {\alpha'}_k}
P^k_{\alpha_k}$, and
$\sum_{\alpha_k} P^k_{\alpha_k} = I$.  A 
{\it history} $\tilde \alpha$ corresponds to a sequence of outcomes
$\tilde \alpha  = \alpha_1 \ldots \alpha_n$.
The decoherence functional
$D(\tilde \alpha , \tilde \alpha')$ for initial state $\rho_i$ 
is defined to be
$$D(\tilde \alpha, \tilde \alpha') =  {\rm tr}
P^{n}_{\alpha_n} \ldots P^1_{\alpha_1} \rho_i
P^1_{{\alpha'}_1} \ldots P^n_{{\alpha'}_n} = 
{\rm tr}  P^\dagger_{\tilde \alpha} \rho_i P^\dagger_{\tilde \alpha'}
.\eqno(1)$$
If one performs this sequence of measurements, then
the probability for the sequence of events that constitutes
the history $\tilde \alpha$
are given by the diagonal term of the decoherence functional
$p( \tilde \alpha) = D(\tilde \alpha, \tilde \alpha)$.
Note that these probabilities are non-negative and sum to 1.

The off-diagonal terms in the decoherence function measure the degree
of quantum interference between different histories.    When these
terms are comparable to the corresponding on-diagonal terms, it means
that measurements that correspond to earlier events in the history
have a strong effect on the probabilities for later events [35].
For example, in the double-slit experiment, the histories that correspond
to which slit the particle goes through are coherent: consequently,
a measurement that determines through which slit a particle passes
has a strong effect on probability of where the particle lands on
the screen.  Indeed, such a measurement destroys the characteristic
interference pattern of the double-slit experiment.
The presence of coherence means that the probabilities $p(\tilde\alpha)$
fail to obey probability sum rules.   For example, let $P^1_{1}, P^1_2$
project onto the states that go through slit $1$ or slit $2$ at
time $t_1$, and $P^2_{x}$ project onto states that land on the
screen at point $x$.   The presence of coherence means that
$${\rm tr} P^2_x \rho_i = \sum_{jj'}
{\rm tr}
P^2_x P^1_j\rho P^1_{j'} \neq \sum_j {\rm tr}
P^2_x P^1_j\rho P^1_j.\eqno(2)$$
That is, the probabilities for where the particle lands on the screen
in the absence of a measurement of which slit it passed through are 
different from the probabilities for where the particle lands on the
screen given that a measurement has been made.
More generally, if
$$|D(\tilde \alpha, \tilde \alpha')|^2 / D(\tilde \alpha, \tilde \alpha)
D(\tilde \alpha', \tilde \alpha') \leq \epsilon << 1, \eqno(3)$$
then the histories are said to be approximately decoherent:
such histories obey probability sum rules to accuracy $\epsilon$.

The decoherence functional characterizes the degree to which measurements
affect the future behavior of a quantum system.   If a system decoheres
with respect to a particular set of measurements, then the measurements
made in the past have minimal effect on the outcomes of measurements made
in the future.   If this is so, then we can assign probabilities to 
the sequence of events corresponding to the measurement outcomes {\it
whether the measurements are actually performed or not.}   That is,
although they are defined in terms of the mathematical apparatus of
measurements -- projection operators or more generally positive operator
valued measures (POVMs) -- decoherent histories represent a method for
assigning probabilities to sequences of events in the absence of measurement.
In the words of Griffiths [1-3], 
decoherent histories refer to sequences of events
`that we can talk about at the breakfast table.'   In the double
slit experiment, we are not allowed to talk about the particle going
through either one slit or the other, because to explain the interference
pattern on the screen, it must go through both at once.   

We now apply the method of decoherent histories to show that
stationary states can exhibit non-trivial time-dependent decoherent histories.

First look at histories that trivially decohere -- the histories for energy
eigenstates of a closed physical system.   In this
case, $P^k_{j} = |E_j\rangle\langle E_j|$ for energy
eigenstates $|E_j\rangle$.   Note $P^k_j$ is independent of the time
step $k$. We have 
$P^k_{j} P^k_{j'} = 
\delta_{jj'} P_j$. 
and $D(\tilde j; \tilde j') \propto \delta_{\tilde j
\tilde j'}$, independent of the initial state
$\rho_i$.    Here the initial projection takes $\rho_i$ to
an energy eigenstate, and subsequent projections simply confirm
that the system remains in that state: histories of energy eigenstates
do not exhibit time-dependent fluctuations.

Histories of other variables do exhibit fluctuations, however. 
A system can possess complementary sets of
decoherent histories.  In the harmonic oscillator, for
example, even though energy does not fluctuate, phase does.  
Consider an harmonic oscillator with Hamiltonian
$\hbar\omega a^\dagger a = \hbar\omega \sum_{\ell = 0}^\infty
 |\ell\rangle\langle \ell|$,
where $|\ell\rangle$ is the $\ell$th energy eigenstate.  For simplicity,
restrict attention to the subspace ${\cal H}_N$ of
states whose energy is less than $N\hbar\omega$.
Within this space, we can define phase states $|\phi\rangle
=N^{-1/2} \sum_{\ell=0}^{N-1} e^{i\ell\phi} |\ell\rangle$.  The
phase states evolve in time as $|\phi\rangle \rightarrow
|\phi + \omega t\rangle$.  The
$N$ states $|\phi_j\rangle$ where $\phi_k = 2\pi j/N$ form
an orthonormal basis for ${\cal H}_N$.   Over time $\Delta t
= 2\pi /N\omega$, we have 
$$|\phi_j\rangle \rightarrow 
U_{\Delta t} |\phi_j\rangle =
|\phi_{j+1}\rangle, \eqno(4)$$ 
where $U_{\Delta t} = e^{-iH\Delta t/\hbar}$ and
$j+1$ is defined modulo $N$.
That is, over time $\Delta t$, the states $|\phi_j\rangle$
evolve deterministically into each other.

Suppose that the oscillator starts out in its ground state $|0\rangle$,
and consider the histories defined by measurement operators
$P_j = |\phi_j\rangle \langle \phi_j|$ spaced at intervals $\Delta t$.  
The decoherence functional is
$$D(j_1\ldots j_n,j'_1\ldots j'_n)
= {\rm tr} P_{j_n} U_{\Delta t} \ldots 
U_{\Delta t} P_{j_1} |0\rangle\langle 0| P_{j'_1} U^\dagger_{\Delta t}
\ldots U^\dagger_{\Delta t} P_{j'_n}.\eqno(5)$$
Because of the deterministic evolution of the phase states, we
have $D(j_1\ldots j_n,j'_1\ldots j'_n) = 0$ unless
$j'_n = j_n$, $j_k=j_{k-1} + 1$, and $j'_k = j'_{k-1} + 1$.
That is, the off-diagonal terms of 
$D(j_1\ldots j_n,j'_1\ldots j'_n)$ are all zero, and the on-diagonal
terms reflect the deterministic nature of the time evolution.
The first projection yields equal probabilities $1/N$ for all phase
states $|\phi_k\rangle$, and the subsequent evolution is entirely
deterministic.   The phase measurement corresponds to decoherent
histories, even though the initial state is the ground state.

The set of decoherent histories corresponding to phase state
evolution describes a quite different type of behavior from the
behavior given by the set of decoherent histories
corresponding to energy eigenstates.  Described in terms of
energy, the system remains static.   Described in terms of
phase, the system fluctuates.  The two types of histories,
energy and phase, represent complementary ways of describing the 
evolution of the same physical system.  

As one might expect, histories that mix phase and energy eigenstates
fail to decohere.   Indeed, histories that begin in
an energy eigenstate, progresses through a sequence of phase states,
and then end in an energy eigenstate, are fully coherent: it is
straightforward to show that the off-diagonal parts of the 
decoherence functional are of the same size as the on-diagonal parts. 
In particular, if one starts in the ground state, and then ends
in the ground state, phase fluctuations do not decohere.   
In general, histories that begin
and end in a pure state decohere only if the histories are
completely deterministic.   When the final projector is a pure 
state, $P_n = |\phi\rangle\langle\phi|$, we have
$$D(\tilde \alpha; \tilde \alpha')
= \langle \phi | P_{\tilde\alpha} |\psi\rangle\langle \psi| 
P^\dagger_{\tilde\alpha'} |\phi\rangle, \eqno(6)$$
and
$$|D(\tilde \alpha, \tilde \alpha')|^2 / D(\tilde \alpha, \tilde \alpha)
D(\tilde \alpha', \tilde \alpha')=1,\eqno(7)$$
unless $D(\tilde \alpha; \tilde \alpha') \propto \delta_{\tilde \alpha
\tilde\alpha'}$: to be decoherent, the histories must be deterministic. 
So time-dependent histories of probabilistic
fluctuations that begin in the vacuum and end in
the vacuum are coherent.

By contrast, histories that begin in a mixed state such as a thermal
state, and end in an mixed state, can be decoherent.
Now consider decoherent histories of thermal states.   Such states
can either arise from interaction with a reservoir at temperature
$T = 1/k_B \beta$, or as subsystems of a larger system that is
in a pure state.   The latter case arises in gravitational contexts
such as Hawking radiation, Unruh radiation, and de Sitter space.
The mathematical question of whether or not histories decohere
depends only on the thermal form of the state and on the dynamics,
not on whether the system is thermal because it is
interacting with a reservoir or thermal because it is entangled.
A considerable literature shows that stationary, thermal states
exhibit non-trivial, temporally fluctuating, decoherent histories 
[12-35].   The positions of particles that begin in thermal
states and that undergo Brownian motion
exhibit decoherent histories, as do hydrodynamic variables -- the
coarse-grained values of quantum field, energies, and particle
densities.    

\bigskip\noindent{\it Decoherent histories and quantum jumps}

When the system in question is an open system
interacting with its environment, or equivalently a subsystem
of a larger system, then the quantum jump picture
yields decoherent histories for sequences of projections
corresponding to the jump operators [18-22].  The decoherence
of histories of quantum jumps 
allows one to relate the decoherent histories approach to the
idea of environmentally induced decoherence.  In particular, 
if the subsystem's dynamics can be described by a Lindblad
equation, then the resulting
stochastic Schr\"odinger equation intrinsically 
gives rise to decoherent histories.  So, for example, a subsystem
in a stationary thermal state that is a fixed point of the Lindblad 
equation undergoes decoherent histories described by probabilistic
jumping from state to state.  Such decoherent histories exist
both when the open system is in a thermal
state because of its interaction with a thermal environment, and when
it is a subsystem of a larger system in a pure state.  The automatic
existence of decoherent histories corresponding to histories of the
stochastic Schr\"odinger equation and of quantum state diffusion is
particularly useful as the jump operators for systems weakly
coupled to a Markovian environment represent jumps between energy
eigenstates.  For such systems, we are allowed to talk at the breakfast table
about the system hopping thermally from energy eigenstate to
energy eigenstate as described by the Bloch-Redfield equation,
even though the system as a whole is a stationary thermal state.

We present here a simple derivation of why 
systems that evolve according to a Lindblad equation exhibit decoherent
histories.  
In contrast to previous derivations [18-22], which focus
on an Itoh calculus derivation of the relation between quantum
jumps and decoherent histories, the derivation given here
is  based on environmentally induced decoherence.
The Lindblad equation represents the most general 
infinitesimal completely positive (i.e., legal)
time evolution for a quantum system.
A general Lindblad equation takes the form 
$${\partial \rho\over \partial t} = -i[H,\rho] - {\gamma/2} \sum_j \big(
L_j^\dagger L_j \rho - 2L_j\rho L_j^\dagger + \rho L_j^\dagger L_j\big).
\eqno(8)$$
The Lindblad equation for a given system interacting with its environment
can be derived by starting with system and environment in the uncorrelated
state $\rho_S\otimes \rho_E$, and applying the unitary system-environment
time evolution $U(\Delta t)$ over a time $\Delta t$ equal to the correlation
time of the environment.  The system evolves to  
$$\rho_S(0) \rightarrow \rho_S(\Delta t) = {\rm tr}_E
 \rho U(\Delta t) \rho_S\otimes \rho_E U^\dagger(\Delta t). \eqno(9)$$
Expanding to second order in $\Delta t$ yields the infinitesimal form
of the Lindblad equation (8).  That is, in addition to being the
general infinitesimal form for a completely positive map, the Lindblad
equation has a physical interpretation as an approximate infinitesimal
time evolution for a system interacting unitarily with an environment
whose correlations decay over a characteristic time.
 
To look at decoherent histories,
for simplicity consider the case where there is only one Lindblad
operator $L_1 = L$ and $H=0$: 
$\partial\rho/\partial t = (-\gamma/2) ( L^\dagger L \rho
-2 L\rho L^\dagger +\rho L^\dagger L)$.  
Use the polar decomposition
to write $L = U A$, where $U$ is unitary, $U^\dagger= U^{-1}$, and $A$ is
Hermitian, $A=A^\dagger$.  The infinitesimal
dynamics generated by the Lindblad equation over time $\Delta t$
is equivalent to 
the following measurement plus feedback procedure:

\bigskip\noindent  (1) Make a generalized measurement on the
system with POVM operators $ M_1= A^2 \gamma \Delta t$ and
$M_2 = 1- A^2  \gamma\Delta t$.   With probability 
$ p_1=  \gamma \Delta t ~ {\rm tr} 
A^2 \rho $ the system goes to the state
$ \rho_1 = (1/p_1) A \rho A$, and with probability $ p_0 = 1-p_1$ the system
goes to the state 
$ \rho_0 = (1/p_0)\sqrt{1- A^2 \gamma \Delta t} ~ \rho ~ 
\sqrt{1- A^2 \gamma \Delta t} \approx \rho_0$.   
Because any generalized measurement can be written as a von Neumann
measurement on system plus an ancilla [55], we have
$\rho_1 = {p_1}^{-1} P_1 \rho\otimes \sigma P_1$,
$\rho_0 = {p_0}^{-1} P_0 \rho \otimes \sigma P_0$,
for projectors $P_1$, $P_0 = 1-P_1$, and ancilla in state $\sigma$.
(This technique shows how to generalize decoherent histories
from projective measurements to generalized measurements [35].)

\bigskip\noindent (2) Now feed back the result of the
measurement.  If the result of the measurement is the
state 1, apply the unitary transformation $U$.   If the result is
0, do nothing.   The system is now in the state
$$ \eqalign{
\rho' &= \sqrt{1- A^2  \gamma\Delta t} ~ \rho ~ \sqrt{1- A^2  \gamma\Delta t} 
+ L\rho L^\dagger \gamma  \Delta t \cr
&= \rho -( \gamma \Delta t/2)  
( L^\dagger L \rho-2 L\rho L^\dagger +\rho L^\dagger L)   
+ O(\Delta t^2). \cr} \eqno(9)$$

\bigskip\noindent
Because the Lindblad equation is mathematically equivalent to 
projective measurement on system plus ancilla followed by
unitary feedback, the set of histories of the system plus ancilla corresponding
to the projections $P_0,P_1$ repeated at time intervals $\Delta t$
are decoherent.  (Note that in this picture the unitary time evolutions
between projections depend on the previous history of projections, 
corresponding to the more general model of decoherent histories given
by Gell-Mann and Hartle [7-15].)  
One can think of this demonstration of decoherence
as a {\it derivation} of the stochastic Schr\"odinger equation or of
a quantum jump model.  Because the histories are decoherent,
we can describe the time evolution of the system in terms of
a stochastic process: over time $\Delta t$ the
system goes to the state $\rho_1 
= {p_1}^{-1} L\rho L^\dagger$ with probability $p_1 =
{\rm tr} L\rho L^\dagger \gamma \Delta t$, or remains
in the state 
$\rho_0 = 
\sqrt{1- A^2  \gamma\Delta t} ~ \rho ~ \sqrt{1- A^2  \gamma\Delta t}$
with probability $p_0 = 1-p_1$.
The treatment of the general Lindblad equation is essentially
the same, except now there are multiple types of quantum jumps
that can occur, one type for each Lindblad operator, and the
time evolution in between jumps includes the effect
of the system Hamiltonian.
Note that the histories induced by the Lindblad equation
are decoherent for any initial state, including stationary states
such as thermal states or energy eigenstates.  

This result establishes that an open system whose time evolution
is governed by a Lindblad equation can be thought of as undergoing
quantum jumps even when it is in a stationary state.  The histories
corresponding to different sequences of quantum jumps are decoherent.   

As an example, consider a harmonic oscillator with Hamiltonian
$H = \hbar\omega \sum_\ell |\ell\rangle\langle \ell| 
= \hbar \omega a^\dagger a$ as above,
interacting linearly with
a bath of modes of the electromagnetic field at temperature $T = 1/\beta$.
The oscillator obeys the Lindblad equation
$${\partial \rho\over \partial t}
= -i [ H,\rho] - {\gamma_+\over 2}\big( a a^\dagger \rho
- 2 a^\dagger \rho a + \rho a a^\dagger \big) 
- {\gamma_-\over 2}\big( a^\dagger a \rho
- 2 a\rho a^\dagger + \rho  a^\dagger a \big), \eqno(10)$$
where $\gamma_+/\gamma_- = e^{-\beta \omega}$.  The thermal state
$\rho_{th} = (1/Z) e^{-\beta H}$ is a stationary state of the
time evolution.  In the thermal state, the oscillator exhibits
decoherent histories over sequences of energy eigenstates,
in which the $n$th energy eigenstates absorbs photons at a rate
$n\gamma_+$ and emits photons at a rate $n\gamma_-$.

\bigskip\noindent{\it Decoherent histories for open systems over times
longer than the relaxation time}

Consider an open system that relaxes over time $\tau$ to a fixed
state $\rho_0$ of the system's Hamiltonian $H$, e.g., a thermal state
$\rho_0 = (1/Z) e^{-\beta H}$.  Suppose that the time intervals between the
projectors $P_{\alpha_j}$ are much longer than $\tau$.   Then {\it any}
set of histories decoheres (i.e., not merely the jump histories given
by the Lindblad operators).   The reason is simple: because the system
relaxes to the same state independent of the input state, if one waits
for $>> \tau$, the probabilities for measurement are just given by the
probabilities for measurement on $\rho_0$, independent of whether
some measurement was made long ago or not.

\bigskip\noindent{\it Time-dependent fluctuations in stationary states}

The notion that a stationary quantum state does not fluctuate in time seems
at first a perfectly reasonable one [54].   However, the time-independent
history of a stationary state can also be decomposed as a quantum
superposition of time-dependent fluctuating histories.   Under a wide
variety of circumstances, those histories decohere, and so we
are free to describe the time evolution of such systems in 
terms of those histories.   A recent paper [54] suggested
the contrary, it is worth discussing briefly the
decoherent histories in thermal states, why do Boddy {\it et al.} 
decide that decoherent histories are not possible in such states [54]?
There are two reasons.   First, they use a time-symmetric version of
decoherent histories that includes both initial and final states.  
Second, they use decoherent histories with only one intermediate
set of events between those initial and final states.   While
it is true that such histories do not decohere, it is unclear why
one should to restrict one's attention so such histories.

We review their argument.
The time-symmetric version of decoherent histories is appropriate
when all or part of the universe possesses a final state, as in
spatially and temporally compact universes whose state is computed
by the Hartle-Hawking imaginary time procedure, or in the Horowitz-Maldacena
model of black hole evaporation.  It does not seem that such
a situation holds in inflationary models, and so it is unclear
why this formalism should be applied here.   

The decoherence functional
$D(\tilde \alpha , \tilde \alpha')$ for initial state $\rho_i$
and final state $\rho_f$ is defined to be
$$D(\tilde \alpha, \tilde \alpha') = Z^{-1}  {\rm tr} \rho_f
P^{n}_{\alpha_n} \ldots P^1_{\alpha_1} \rho_i
P^1_{{\alpha'}_1} \ldots P^n_{{\alpha'}_n} = Z^{-1}
{\rm tr} \rho_f P^\dagger_{\tilde \alpha} \rho_i P_{\tilde \alpha'}
,\eqno(11)$$
where $Z^{-1} = {\rm tr} \rho_f \rho_i$.  As before, histories
decohere if the off-diagonal terms in the decoherence functional
are small compared with the on-diagonal ones, given that initial
state is $\rho_i$ and the final state is $\rho_f$. 
The addition of
the final state in the decoherence function is equivalent to adding
one additional measurement operator, whose final projection correponds
to a measurement revealing that the system is in $\rho_f$.  Such a
measurement can be performed, for example, by adjoining an ancillary
system in state $\rho_A$ and performing a projection $P$ on system
and ancilla such that ${\rm tr}_A P I\otimes \rho_A P = \rho_f$.

Boddy {\it et al.} consider closed quantum systems and
investigate situations where the initial and final state are both
stationary states of the system dynamics.  They consider
histories with a single projection between the initial and
final state, for which the decoherence functional is
$$ Z^{-1} {\rm tr} \rho_f P_\alpha \rho_i P_{\alpha'}.\eqno(12)$$
Boddy {\it et al.} note correctly that this decoherence
functional is not dependent on the evolution times between
the initial state, the projections, and the final state.
They conclude (also correctly), that such histories do not
exhibit perfect decoherence.   

Two questions: first,
why use the time-symmetric version of the
decoherent histories formalism?   Boddy {\it et al.}'s justification is that
they are interested in histories that begin and end in a thermal
state, e.g., the state of the fields de Sitter space.  When looking
at cosmological histories, however, there is no particular reason
for making this restriction unless one desires artificially to 
restrict the set of possible decoherent histories.   If one
uses the ordinary formulation of decoherent histories, starting
from initial states and evolving forward in time, thermal
states such as those in de Sitter space can exhibit a wide variety
of non-trivial decoherent histories.   The second question is simpler:
why restrict attention to histories with only one set of events?
Suppose that one adds a second set of events, so that the decoherence
functional is
$$ Z^{-1} {\rm tr} \rho_f P_{\alpha_2} P_{\alpha_1} 
\rho_i P_{{\alpha'}_1}  P_{{\alpha'}_2}.\eqno(13)$$
In this case, because the projection operators in the Heisenberg picture
do depend on time, the decoherence functional depends on the
time difference between the two sets of events.   As noted above,
decoherent histories over multiple sets of events also exhibit
time dependence even when the initial and final states are stationary.   

Boddy {\it et al.} are correct that the particular 
set of histories that they investigate do not decohere.
It would be a mistake to conclude, however
that systems that begin and end in
stationary states cannot exhibit decoherent histories.   
Indeed, a rather trivial counterexample occurs when both
$\rho_i$ and $\rho_f$ are the fully mixed stationary state
$I/d$, corresponding to a thermal state with infinite temperature.
In this case, the different sets of histories described above for
both open and for closed systems naturally decohere.   When
$\rho_i$ and $\rho_f$ are thermal states $Z^{-1}e^{-H/kT}$
at finite temperature, then we can project those states
onto a typical subspace of dimension $d \propto e^S$,
where $S = H/kT - \ln Z$ is the entropy of the thermal state.
Once again, these states naturally decohere as in the examples
above.

\bigskip\noindent{\it Quantum cosmology and decoherent histories}

Having established that stationary states do indeed exhibit quantum
fluctuations, at least by the criterion of decoherent histories,
let's turn to quantum cosmology in models of eternal inflation.  
In such models, vacuum energy induces an effective cosmological
term $\Lambda$, causing spacetime locally to resemble de Sitter space.
Inertial observers witness an event horizon at distance 
$\ell = \sqrt{3/\Lambda}$,
and detect horizon radiation in a thermal state with temperature
$T=1/2\pi\ell$.   

Look at decoherent histories corresponding to measurements made
by such an inertial observer as the universe undergoes inflation and
settles down into de Sitter space with a small cosmological term; after
a long time (estimated below) the observer
encounters a local thermal fluctuation that gives rise
to a region that possesses a high cosmological term and that is large
enough to seed another inflationary epoch.  Actually to make these measurements,
such an inertial observer would have to be a hardy individual, 
capable of surviving extreme temperatures and curvatures.   
(Assume, however, that the observer does not fall into a black hole.)
The whole point of decoherent histories, however, is to be
able to assign probabilities to events whether or not the
measurements corresponding to the events are actually performed.
An hypothetical inertial observer suffices to assign decoherent
histories.   The measurements correspond to coarse grained
observations of fields, energy densities, pressure, etc., in
the local vicinity of the observer. 
As noted above in [10-35], 
such histories generically exhibit approximate decoherence.

Consider decoherent histories that correspond to the inertial
observer making coarse-grained measurements of the fields in her vicinity
together with the effective cosmological term $\Lambda$.   
The events along any history depend on the value of $\Lambda_0$
in the first projection in the history.   Suppose for the moment 
that $\Lambda_0$ is large,
corresponding to high energy in the false vacuum.  The observer
initially sees thermal fluctuations in the fields
corresponding to de Sitter space at high temperature, and witnesses
the surrounding spacetime inflate at a rate $\propto \sqrt \Lambda_0$.      
After a characteristic time-scale $\tau_0$,
the observer enters a region in which false vacuum decays.   
That is, in such a history
the universe undergoes inflation via the usual scenarios, yielding
a universe more or less like our own.
If at the end of inflation the cosmological term is non-zero, then the region
in the vicinity of the observer will eventually settle down to de Sitter
space again with effective cosmological term $\Lambda_1 < \Lambda_0$ and
a horizon at distance $\ell_1 = \sqrt{3/\Lambda_1}$.  The
state within the horizon is thermal with temperature $T_1
= 1/2\pi\ell_1$.  As above,
this thermal state can exhibit decoherent histories corresponding
to fluctuations in local energy density.   Eventually, the region
in the vicinity of the observer will exhibit a fluctuation that
takes it back to a regime with high $\Lambda$ and will begin
inflating again.   

To estimate the time $\tau_1$ it takes for rapid inflation
to recommence, we look at how long it takes for a thermal fluctuation
to generate a reinflating region.   By assumption, the dynamics
possesses at least one quantum state for a region of radius 
$\ell=C\ell_0 = C \sqrt{3/\Lambda_0}$ 
that undergoes inflation with effective cosmological
term $\Lambda_0$, where $C$ is a positive $O(1)$ constant.     
The energy that has to be collected from the thermal
radiation in de Sitter space to attain energy density $\Lambda_0$ over
a volume $\ell^3$ is 
$$\Delta E =  \ell^3 \Lambda_0 =     
3C^3\ell_0.\eqno(14)$$   
The de Sitter radiation has 
temperature $T_1 = 1/2\pi\ell_1$.
To collect the energy $\Delta E$ within a region of spatial
extent $\ell$ reduces the entropy of the surrounding de Sitter
radiation by $\Delta S =  \Delta E/T_1$.   The probability that a thermal
fluctuation at the de Sitter temperature gives rise to
a region with the energy density needed to reinflate is thus
$e^{-\Delta S} = e^{-6\pi C^3 \ell_0 \ell_1}$.

It is not enough for the energy required for reinflation to assemble
itself: the fields must also be in the proper false vacuum state.   
If the energy
is assembled in a random state, the overlap with the proper inflating
state goes as $e^{-S_0}$, where $S_0 = \pi \ell_0^2$ is the de Sitter
entropy for the fields in the region -- i.e., the maximum entropy for
the region.   The overall thermal
probability for a fluctuation that creates the inflating region then 
goes as
$$ e^{-\Delta E/T_1 - S_0}
= e^{-6\pi C^3 \ell_0 \ell_1 - \pi \ell_0^2} 
= e^{-6C^3 \sqrt{ S_0 S_1} - S_0} .\eqno(15)$$ 
Here $S_0 = \pi \ell_0^2$ is the entropy of the high energy
density de Sitter space with cosmological term $\Lambda_0$, and $S_1$
is the entropy of the low energy density de Sitter space.
Remarkably -- given the simple and non-gravitational
nature of the argument --  equation (15) reproduces (up to the
value of the constant $C$)
the Farhi-Guth-Guven formula [56] for the thermal probability of 
exciting an inflating volume of spatial extent $ \approx \ell_0$.   
Equation (15) shows that the decoherent histories
corresponding to a hardy inertial observer reproduce the eternal
inflation picture suggested by Albrecht [49].

Equation (15) gives the thermal probability for a fluctuation
that can re-ignite inflation.   To estimate the time it takes
for such a fluctuation to arise, note that in de Sitter space
with cosmological term $\Lambda_1$,
the characteristic time
for fluctuations to arise and decay is $\approx\pi/T 
= 4\pi^2 \ell_1$, yielding a time       
$$ \tau_1 \approx 
4\pi^2 \ell_1 e^{ (6\pi C^3 \ell_0 \ell_1 + \pi \ell_0^2)},\eqno(16)$$
for the inertial observer to encounter another rapidly inflating region.
Note that process of reinflation is much more likely to begin with a small
region with large cosmological term, than with a large region with small 
cosmological term.   This observation suggests that reinflation should 
typically begin at a scale close to the Planck scale. 

\bigskip\noindent{\it Initial state}

As in [49] this argument yields an ergodic model of eternal inflation.
Inflation takes place at high energy scales; the value of the field
rolls downhill, yielding an FRW
universe with the usual features; eventually, the presence of a small
cosmological term yields a fluctuation that causes inflation to begin
again with large cosmological term.  By the arguments given above,
such histories are generically decoherent.   
The initial state of the universe as a whole can be taken to be the
stationary state given by the ergodic average of the the state of
the universe over time.  

Note that, although ergodic, the histories seen by an inertial observer
are not time-reversal invariant.   Every time rapid inflation
begins again at high $\Lambda_0$, it 
supplies a large source of free energy so that 
the inertial observer sees entropy increasing, consistent with the second 
law of thermodynamics.    This time-asymmetry of the individual
histories is nonetheless consistent with time-reversal invariance 
of the initial, stationary state of the universe {\it and} of its dynamics.
If instead of using the theory of decoherent histories with an
initial state,  we use decoherent histories that end in a final state, 
then we can decompose the stationary state of the universe into a 
superposition of decoherent histories that end in a final state
of high $\Lambda_0$, and that evolve backward in time.   The histories
that correspond to inertial observers with a fixed final state
are then just the time reversed version of histories
with a fixed initial state: even though they are moving in the
opposite direction in time, the time-reversed inertial observers
still observe entropy increasing.

\bigskip\noindent{\it Re-inflation versus Boltzmann brains}

Now compare the probability of the inertial observer encountering
another rapidly inflating region with the probability
of encountering a thermal fluctuation that mimics some small
piece of our universe, e.g., a `Boltzmann brain.'   The argument
for the probability of recreating an inflating region via
a thermal fluctuation is readily generalized to calculating
the probability of recreating any system with energy $\Delta E$ and
entropy $S = S_{max} - \Delta S$, where $S_{max}$ is the maximum
entropy for the system confined to the volume in which it is
created.   The thermal probability of such a fluctuation is
$e^{-\Delta E/T_1 - \Delta S}$.   
Comparing with equation (14) we see that as long as the energy
$\Delta E$ in the Boltzmann brain is greater than the energy
$\approx \ell_0$ required to reignite inflation, then the 
inertial observer is more likely to encounter brains that
arise by the ordinary process of evolution in an FRW universe
rather than ones that arise from thermal fluctuations.
If $\ell_0$ is at the grand unification scale or a shorter
length scale, e.g. the Planck scale,  then the vast majority 
of brains encountered by the observer over its infinite
history will be the usual kind of brains. 

\bigskip\noindent {\it Summary}

This paper investigated the question of whether stationary states
can exhibit non-trivial temporal fluctuations.   Viewed through
the lens of decoherent histories, the answer is an unqualified Yes.
Closed systems in energy eigenstates exhibit decoherent histories
with non-trivial temporal fluctuations.  Open systems that are
subsystems of larger systems exhibit decoherent histories that
correspond to the quantum jump or stochastic Schr\"odinger model.
We then investigated decoherent histories that correspond to
observations made by an inertial observer in models
of eternal inflation.   Such histories contain periods
of inflation leading to FRW universes with small cosmological
term; thermal fluctuations from the cosmological term then
reignite inflation.   The time scale required to reignite
inflation is long compared with the horizon scale but much
shorter than the time required to generate thermal fluctuations
that mimic systems that evolved from initial low-entropy states.

\vfill
\noindent{\it Acknowledgements:} This work was supported by ARO, DARPA,
and AFOSR.   The author thanks Andy Albrecht, Alan Guth, Allan Adams, 
Sean Carroll, Jonathan Halliwell, Jim Hartle, Don Page, and Dan Roberts
for extensive discussions.

\vfil\eject
\noindent{\it References}

\vskip 1cm

\noindent [1] R.B. Griffiths,  {\it J. Stat. Phys.} {\bf 36}, 219 (1984). 

\smallskip\noindent [2] R.B. Griffiths, {\it Phys. Rev. Lett.} {\bf 70}, i
2201 (1993).

\smallskip\noindent [3] R.B. Griffiths, {\it Phys. Rev. A} {\bf 54},
 2759 (1996); {\it Phys. Rev. A} {\bf 57}, 1604 (1998).

\smallskip\noindent [4] R. Omn\`es, {\it J. Stat. Phys. } {\bf 53}, 893 (1988); 
{\it J. Stat. Phys.} {\bf 53}, 933 (1988); 
{\it J. Stat. Phys.} {\bf 53}, 957 (1988); {\it J. Stat. Phys.} {\bf 57}, 
357 (1989);  

\smallskip\noindent [5]  R. Omn\`es, {\it Ann. Phys.} {\bf 201}, 354 (1990) 

\smallskip\noindent [6]  R. Omn\`es {\it Rev. Mod. Phys.} {\bf 64}, 339 (1992).

\smallskip\noindent [7] R. Omn\`es, {\it J. Math. Phys.} {\bf 38}, 697 (1997)

\smallskip\noindent [8] M. Gell-Mann, J.B. Hartle, 
in {\it Complexity, Entropy and the Physics of Information,} SFI
Studies in the Sciences of Complexity, Vol. VIII, 
W.H. Zurek, ed., (Addison Wesley, Reading, 1990); 

\smallskip\noindent [9] M. Gell-Mann, J.B. Hartle, in
{\it Proceedings of the Third International Symposium on the 
Foundations of Quantum Mechanics in the Light of New Technology,}
S. Kobayashi, H. Ezawa, Y. Murayama and S. Nomura, eds. 
(Physical Society of Japan, Tokyo, 1990).

\smallskip\noindent [10] M.Gell-Mann and J.B.Hartle, 
{\it Phys. Rev. D} {\bf 47}, 3345 (1993).

\smallskip\noindent [11] M. Gell-Mann, J.B. Hartle, 
in {\it Physical Origins of Time Asymmetry,}
J. Halliwell, J. Perez-Mercader, and W. Zurek, eds.,
Cambridge University Press, Cambridge (1994), arXiv: gr-qc/9304023.

\smallskip\noindent [12] J.B.Hartle, in 
{\it Proceedings of the Cornelius Lanczos International Centenary 
Confererence,} J.D. Brown, M.T. Chu, D.C. Ellison and R.J.Plemmons, eds.,
(SIAM, Philadelphia, 1994); arXiv: gr-qc/9404017.

\smallskip\noindent [13] J.B. Hartle, in 
{\it Proceedings of the 1992 Les Houches School, Gravity and 
its Quantizations,} B. Julia, J. Zinn-Justin, eds. 
(Elsevier Science B.V., 1995).

\smallskip\noindent [14] J.B. Hartle, D. Marolf, 
{\it Phys. Rev. D} {\bf 56}, 6247 (1997).

\smallskip\noindent [15] M. Gell-Mann, J.B. Hartle, 
{\it Phys. Rev. A} {\bf 85}, 062120 (2012).

\smallskip\noindent [16] J.J. Halliwell, {\it Phys. Rev. D} {\bf 39}, 
2912 (1989).

\smallskip\noindent [17] H.F. Dowker, J.J.Halliwell, 
{\it Phys. Rev. D} {\bf 46}, 1580 (1992).

\smallskip\noindent [18] L. Di\'osi, N. Gisin, J. Halliwell, I.C. Percival,
{\it Phys. Rev. Lett.} {\bf 74}, 203 (1995);
arXiv: gr-qc/9403047.

\smallskip\noindent [19] J.J. Halliwell, A. Zoupas, 
{\it Phys. Rev. D} {\bf 52}, 7294 (1995); 
arXiv: quant-ph/9503008.

\smallskip\noindent [20] J.J. Halliwell, 
{\it Ann. NY Acad. Sci.} {\bf 755}, 726–740 (1995); arXiv: gr-qc/9304039.

\smallskip\noindent [21] T.A. Brun, J.J. Halliwell, 
{\it Phys. Rev. D} {\bf 54} 2899-2912 (1996); arXiv:quant-ph/9601004.

\smallskip\noindent [22] T.A. Brun, {\it Phys. Rev. Lett.} {\bf 78}
1833-1837 (1997); arXiv:quant-ph/9606025.

\smallskip\noindent [23] J.J. Halliwell, {\it Phys. Rev. D} {\bf 58}
105015 (1998);  arXiv:quant-ph/9805062.

\smallskip\noindent [24] J.J. Halliwell, {\it Phys. Rev. Lett.} {\bf 83}
2481 (1999); arXiv: quant-ph/9905094.

\smallskip\noindent [25] J.J. Halliwell, J. Thorwart, {\it Phys. Rev. D} 
{\bf 64}, 124018 (2001). 

\smallskip\noindent [26] J.J. Halliwell, J.Thorwart, {\it Phys. Rev. D} 
{\bf 65}, 104009 (2002). 

\smallskip\noindent [27] J.J. Halliwell, {\it Phys Rev D}
{\bf 68}, 025018 (2003); arXiv: quant-ph/0305084.

\smallskip\noindent [28] J.J. Halliwell, {\it Contemp. Phys.}
{\bf 46}, 93-104 (2005). 

\smallskip\noindent [29] J.J. Halliwell, P. Wallden, 
{\it Phys. Rev. D} {\bf 73}, 024011 (2006).

\smallskip\noindent  [30] J.J. Halliwell, {\it Phys. Rev. D} {\bf 80}, 
124032 (2009).

\smallskip\noindent [31] J.J. Halliwell, {\it Phys. Rev. A} {\bf 72}, 
042109 (2005).

\smallskip\noindent [32]  J.J. Halliwell, {\it  Phys. Rev. D} {\bf 63}
085013 (2001).

\smallskip\noindent  [33] J.J. Halliwell, {\it J. Phys.: Conf. Ser.} 
{\bf 306}, 012023 (2011); arXiv: 1108.5991.

\smallskip\noindent [34] F. Dowker, A. Kent, 
{\it  J. Stat. Phys.} {\bf 82} 1575-1646 (1996);
 arXiv: gr-qc/9412067.

\smallskip\noindent [35] S. Lloyd,
`Decoherent histories and generalized measurements,'
 arXiv: quant-ph/0504155.

\smallskip\noindent [36] L. Boltzmann, {\it Nature}
{\bf 51}, 413 (1895).

\smallskip\noindent [37] A.S. Eddington, (1931), reprinted in 
{\it The Book of the Cosmos: Imagining the Universe from Heraclitus to 
Hawking,} D. R. Danielson, ed., (Perseus, Cambridge, Mass., 2000), p. 406.

\smallskip\noindent [38] R. P. Feynman, The character of physical law (MIT Press Cambridge, 1965) 
p.173. 

\smallskip\noindent [39] D. N. Page, {\it Nature} {\bf 304}, 39 (1983)

\smallskip\noindent [40] J. Barrow and F. Tipler, 
The Anthropic Cosmological Principle (Oxford University Press, Oxford, 1986).

\smallskip\noindent [41] D. Coule, {\it Int. J. Mod. Phys. D} {\bf 12}, 963 
(2003); arXiv: gr-qc/0202104.

\smallskip\noindent [42] A. Albrecht, (2004), in 
{\it Science and Ultimate Reality: From Quantum to Cosmos, 
honoring John Wheeler’s 90th birthday,} 
J.D. Barrow, P.C.W. Davies, C.L. Harper, eds. 
Cambridge University Press (2004); arXiv: astro-ph/0210527.

\smallskip\noindent [43] A. Albrecht, L. Sorbo, {\it Phys. Rev. D} {\bf 70},
063528 (2004); arXiv: hep-th/0410270.

\smallskip\noindent [44] M. Reese, 
{\it Before the Beginning}, p. 221 (Addison-Wesley, 1997).

\smallskip\noindent [45] R. Bousso, B. Freivogel, 
{\it JHEP} {\bf 0706}, 018 (2007); arXiv: hep-th/0610132.

\smallskip\noindent [46] A. Linde, 
{\it J. Cosmol. Astropart. Phys.} {\bf 0701} 022 (2007).

\smallskip\noindent [47] D.N. Page, {\it Phys. Rev. D}
{\bf 78}, 063535, 063536, 2008; arXiv: 
hep-th/0610079, hep-th/0611158.

\smallskip\noindent [48] A. Linde, V. Vanchurin, S. Winitzki,  
{\it JCAP} {\bf 0901}, 031
(2009); arXiv: 0812.0005.

\smallskip\noindent [49] A. Albrecht, {\it J. Phys.: Conf. Ser.}
{\bf 174}, 012006 (2009); arXiv:0906.1047. 

\smallskip\noindent 
[50] A. de Simone, A.H. Guth, A. Linde, M. Noorbala, M.P. Salem, A. Vilenkin,
{\it Phys. Rev. D} {\bf 82}, 063520 (2010); arXiv: 0808.3778.

\smallskip\noindent
[51] A. Linde, M. Noorbala,
{\it JCAP} {\bf 1009}, 008 (2010).

\smallskip\noindent [52]
 A. Albrecht, D. Phillips, `Origin of probabilities and their application 
to the multiverse,' arXiv: 1212.0953.

\smallskip\noindent [53] A. Albrecht, 
`Tuning, Ergodicity, Equilibrium and Cosmology,'
 arXiv: 1401.7309.

\smallskip\noindent [54]
K.K. Boddy, S.M. Carroll, J. Pollack,
`De Sitter Space Without Quantum Fluctuations,'
 arXiv: 1405.0298. 

\smallskip\noindent [55]
M.A. Nielsen, I.L. Chuang, {\it Quantum computation and quantum
information}, Cambridge University Press, Cambridge (2000).

\smallskip\noindent [56]
E. Farhi, A. H. Guth, J. Guven, {\it Nucl. Phys. B} {\bf 339}, 
417–490 (1990).

\vfill\eject\end

\bigskip\noindent{\it Decoherent histories of closed systems}

Now look at decoherent histories of closed systems.   We have already
exhibited two types of pure-state history for the harmonic oscillator,
energy and phase.   Both of these histories decohered because energy
eigenstates evolve into energy eigenstates and phase states evolve into
phase states in a deterministic fashion.   As noted above, such deterministic
histories are the only pure state histories that decohere.   In
general, to decohere, the projectors in the history must be coarse-grained,
projecting onto a large-dimensional subspace of Hilbert space. 
Decoherent histories via the Lindblad equation represent a specific
kind of coarse graining where the projectors act on subsystem:
they take the form $P_S\otimes I_E$, where $P_S$ is a projection operator on
the Hilbert space of the subsystem, and $I_E$ is the identity operator
on the Hilbert space of the environment.  As we will now see,
typical coarse-grained histories for closed systems decohere.   

Consider histories of a closed, $d$-dimensional system
that begins in a pure state $|\psi\rangle$.  Look at histories labeled
by $\tilde \alpha = \alpha_1 \ldots \alpha_n$ as above.   The decoherence
functional takes the form
$$D(\tilde \alpha; \tilde \alpha') =
{\rm tr} P_{\tilde \alpha} |\psi\rangle \langle \psi| P_{{\tilde\alpha}'}
= {1\over \sqrt{ p({\tilde\alpha}) p({\tilde \alpha}')}}
\langle \psi({\tilde\alpha}')|\psi(\alpha)\rangle,\eqno(10)$$ 
where $|\psi(\tilde \alpha)\rangle = (1/\sqrt{ p(\tilde\alpha)})
 P_{\tilde\alpha} |\psi\rangle$, and 
$|\psi({\tilde \alpha}')\rangle = (1/\sqrt{ p{\tilde\alpha}'})
 P_{{\tilde\alpha}'} |\psi\rangle$.   Because $\tilde\alpha$
and ${\tilde \alpha}'$ differ by at least one projection operator,
the generic situation -- true, for example, either for random
Hamiltonians or for randomly selected projections -- is that
$|\psi(\tilde \alpha)\rangle$ have the same overlap as {\it 
random} states in Hilbert space:
$\langle \psi({\tilde\alpha}')|\psi(\alpha)\rangle \approx 1\sqrt d$.
That is, the ratio between off- and on-diagonal terms in
the decoherence functional is typically of the size
$${|D(\tilde \alpha; \tilde \alpha')|^2 \over
D(\tilde \alpha; \tilde \alpha) D(\tilde \alpha'; \tilde \alpha')} =
{|\langle \psi({\tilde\alpha}')|\psi(\alpha)\rangle|^2 \over
p({\tilde\alpha}) p({\tilde \alpha}')} =
{1\over d p({\tilde\alpha}) p({\tilde \alpha}')} \eqno(11).$$
Typical histories decohere until their probabilities are
on the order of one over the square root of the
dimension of the Hilbert space:
after this point, almost all histories become coherent.   In
other words, the maximum amount of information that a set of
decoherent histories can reveal about a system is equal
to the maximum entropy of the system.    

Note that the argument for the decoherence of typical coarse-grained
histories did not depend on initial state: it could be a random
state, the ground state, or a thermal state.   Typical histories will 
still decohere.   

The closed-system and open-system analysis of decoherent histories
are in fact closely related.   The coarse-graining that allows
closed-system histories to decohere splits up the
closed system into `macroscopic' degrees of freedom corresponding
to the labels of the histories, and the remaining `microscopic' degrees of
freedom that are coarse grained over.   The microscopic degrees
of freedom can be thought of as an `environment' that decoheres
the macroscopic degrees of freedom.